# Coexistence of non-trivial van der Waals magnetic orders enable field-free spin-orbit torque switching at room temperature

Bing Zhao[1*], Lakhan Bainsla[1,2], Soheil Ershadrad[3], Lunjie Zeng[4], Roselle Ngaloy[1], Peter Svedlindh[5],

Eva Olsson[4], Biplab Sanyal[3*], Saroj P. Dash[1,6,7*]

[1]*Department of Microtechnology and Nanoscience, Chalmers University of Technology, SE-41296, Göteborg, Sweden.*
[2]*Department of Physics, Indian Institute of Technology Ropar, Roopnagar 140001, India.*
[3]*Department of Physics and Astronomy, Uppsala University, Box-516, 75120 Uppsala, Sweden.*
[4]*Department of Physics, Chalmers University of Technology, 41296, Göteborg, Sweden.*
[5]*Department of Materials Science and Engineering, Uppsala University, Uppsala, SE-751 03 Sweden.*
[6]*Wallenberg Initiative Materials Science for Sustainability, Department of Microtechnology and Nanoscience, Chalmers University of Technology, SE-41296, Göteborg, Sweden.*
[7]*Graphene Center, Chalmers University of Technology, Göteborg, SE-41296 Sweden.*

**Abstract**

The discovery of van der Waals (vdW) magnetic materials exhibiting non-trivial and tunable magnetic interactions can give rise to exotic magnetic states, which are not readily attainable with conventional materials. Such vdW magnets can provide a unique platform for studying new magnetic phenomena and realizing magnetization dynamics for energy-efficient and non-volatile spintronic memory and computing technologies. Here, we discover the coexistence of ferromagnetic and antiferromagnetic orders in vdW magnet $(Co_{0.5}Fe_{0.5})_{5-x}GeTe_2$ (CFGT) CFGT above room temperature, inducing an intrinsic exchange bias and canted perpendicular magnetism. Such non-trivial intrinsic magnetic order enables us to realize energy-efficient, magnetic field-free, and deterministic spin-orbit torque (SOT) switching of CFGT in heterostructure with Pt. The devices show a very large spin Hall conductivity of $3\times10^5$ ℏ/2e $\Omega^{-1}m^{-1}$, a low critical current density of $8\times10^{-2}$ A/µm$^2$, and yield a large SOT effective field of ~8.0 mT per MA/cm$^2$. These experiments, together with density functional theory and Monte Carlo simulations establish coexisting non-trivial magnetic orders in CFGT that enable field-free SOT magnetization dynamics in spintronic devices.

**Introduction**

Exploring new strategies to realize new non-trivial magnetic orders is of great importance for fundamental magnetism, and also for the next generation of memory, logic, communication, and beyond von Neumann computing technologies[1,2]. Such non-trivial magnetic states are required for efficient control of spin-orbit torque (SOT) magnetization dynamics phenomena in spintronic devices for non-volatile information and communication technologies[3–5]. However, for energy-efficient and field-free SOT magnetization switching, the discovery of new materials and methods is needed for enhanced control over the device parameters[2,6–16]. One of the strategies is to utilize heterostructures of ferromagnetic (FM) and antiferromagnetic (AFM) materials with exchange bias to realize field-free SOT magnetization switching[9,10,17,18]. However, there are challenges in fabricating high-quality FM/AFM interfaces with traditional multilayers, and the discovery of a single material with coexisting magnetic orders is necessary.

Recently discovered two-dimensional van der Waals (vdW) magnetic materials have significant potential for such unique non-trivial magnetic states in single material because of their low dimensionality, tunable magnetic anisotropy, proximity-induced interactions, and potential for voltage-controlled magnetism and multi-functionalities of SOT devices[2,19–26]. Furthermore, it is possible to create unique non-trivial spin ordering by controlling the intra- and interlayer magnetic exchange coupling inherent to vdW materials and the magnetic anisotropy depending on atomic position and stacking; however, it is not explored.

Here, we discover non-trivial spin ordering with the coexistence of FM and AFM magnetic ordering in single vdW magnet $(Co_{0.5}Fe_{0.5})_{5-x}GeTe_2$ (CFGT) accompanied by a pronounced exchange bias effect and canted magnetization that persists above room temperature (see Fig.1a), which is not readily attainable with conventional materials. Such non-trivial intrinsic magnetic order enables us to demonstrate an energy-efficient and deterministic field-free SOT-induced magnetization switching of the vdW magnet CFGT at room temperature (see Fig. 1a, inset).

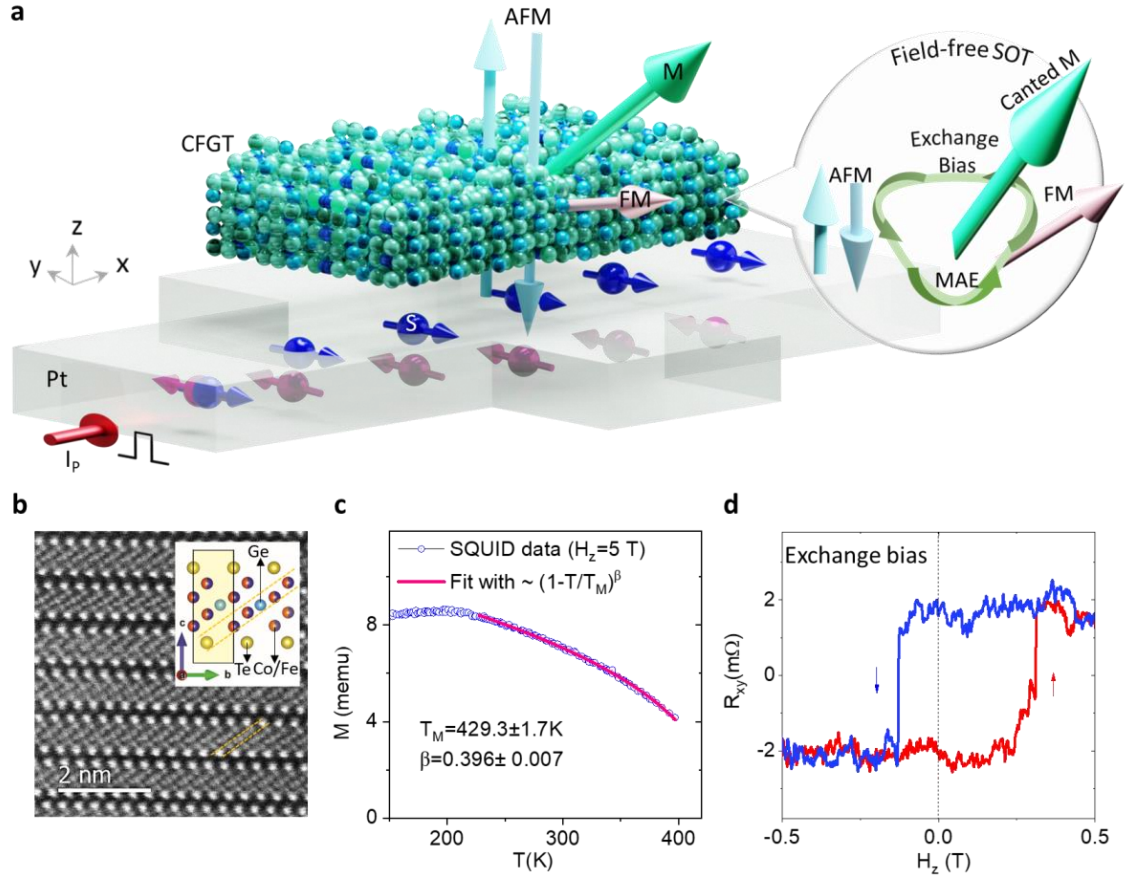

*Figure 1. Coexistence of ferro- and antiferro-magnetic orders in a single atomic stacking nanolayers of $(Co_{0.5}Fe_{0.5})_{5-x}GeTe_2$. **a.** Schematics of CFGT/Pt van der Waals heterostructure used for spin-orbit torque (SOT) experiments. The ferro- (FM) and antiferro-magnetic (AFM) orders in a single atomic stacking nanolayers, creating a net canted magnetization M with exchange bias interaction and magnetocrystalline anisotropy energy (MAE) (in the inset), which is useful for field-free SOT switching. The charge current I applied to Pt generates a spin current $J_s$ with spin polarizations $S_y$ due to the spin Hall effect. **b.** Atomic resolution high angle annular dark field (HAADF) scanning transmission electron microscopy (STEM) images of CFGT. The inset is the atomic model projection along the bc-plane. **c.** Temperature dependence of the magnetization M of bulk CFGT measured with a fixed out-of-plane magnetic field of 5T. The magnetic order temperature $T_M$ is extracted with $\sim(1-T/T_{C(N)})^\beta$ with critical factor $\beta^{27,28}$. **d.** A representative AHE signal at 300 K with clear exchange bias effect ($H_{EB}$=+75 mT) after subtraction of a linear background at a +/-0.6T field range.*

## Results and Discussion

## Coexistence of ferromagnetic and antiferromagnetic orders in CFGT at room temperature: Exchange bias and canted magnetism

The motivation behind using the vdW magnet CFGT is its beyond-room temperature magnetic orders and tunable magnetic properties depending on atomic positions, stacking, and intra- and interlayer magnetic interactions[25,26,29]. CFGT with the wurtzite crystal structure belongs to the P6$_3$mc space group and C$_{6v}$ point group[30]. By substituting around 50% of the Fe atoms with Co, crystals of CFGT alloy were synthesized[26,29,30]. To characterize the atomic structure of CFGT, we performed high-resolution scanning transmission electron microscopy (STEM) measurements as shown in Fig.1b (see details in Methods and Supplementary Note 1 and Supplementary Fig. S1). We observe a very uniform unique AA atomic crystal structure[29,30], with two different types of Fe atomic vacancy positions - an antisymmetric (type I) and a symmetric (type II) crystal structure with $Fe_1$ deficiency sites in the unit cell (Supplementary Fig. S2). Figure 1c shows the temperature-dependent SQUID measurements as a function of fixed out-of-plane field confirms the magnetic behavior of CFGT with a magnetic order temperature $T_M$=429.3±1.7 K.

To investigate the magnetic properties of thin CFGT nanolayers, magnetotransport measurements using the anomalous Hall effect (AHE) were conducted on nanofabricated Hall-bar devices (see Methods section). Figure 1d shows AHE data (Hall resistance $R_{xy}$ as a function of the out-of-plane magnetic field $H_z$) at 300 K, revealing a clear hysteresis loop that confirms the presence of out-of-plane ferromagnetism in CFGT with finite remanence and coercivity at room temperature. More interestingly, a strong exchange bias effect is observed, which is manifested as a horizontal shift in a magnetic hysteresis loop due to the coexistence of intrinsic FM and AFM ordering in CFGT. Our density functional theory (DFT) calculations also show the coexistence of FM and AFM magnetic states. The boxplots in Fig. 2a illustrate these structural dependencies of calculated magnetic ground states, where each dot represents a distinct Co distribution (see details in Supplementary Note 2 and Table S1). The detailed AHE measurements of CFGT nanolayers in both low and high field ranges confirmed the coexistence of FM and AFM states at room temperature below the spin-flip transition, as shown in Supplementary Fig. S3a. We further evaluated the temperature evolution of the FM (AFM) magnetic orders (Supplementary Fig. S3b). The extracted magnetic phase diagram for the CFGT nanolayers as a function of $H_z$ and temperature T is shown in Fig. 2b. In a small field range, the $FM_0$ and AFM orders coexist. A transitional magnetic phase (TFM) appears when the field is large enough to re-align the AFM components. Finally, both AFM and $FM_0$ magnetic phases are aligned with the external field direction to form the ferromagnetic $FM_1$ phase. Based on the evolution of magnitude of AFM(FM) components in the AHE signals (Fig. 2b, inset), we extract the Curie (Néel) temperature $T_C$=360±2 K ($T_N$=395±15 K) of CFGT nanolayers. Furthermore, figures 2c illustrate a more detailed temperature evolution of the AHE signal, where apparent hysteresis is observed in a smaller magnetic field sweep, i.e. with the ferromagnetic FM0 phase. At lower temperatures, the coercivity $H_{c+/-}$ increases, and the exchange bias effect becomes more prominent below 150 K.

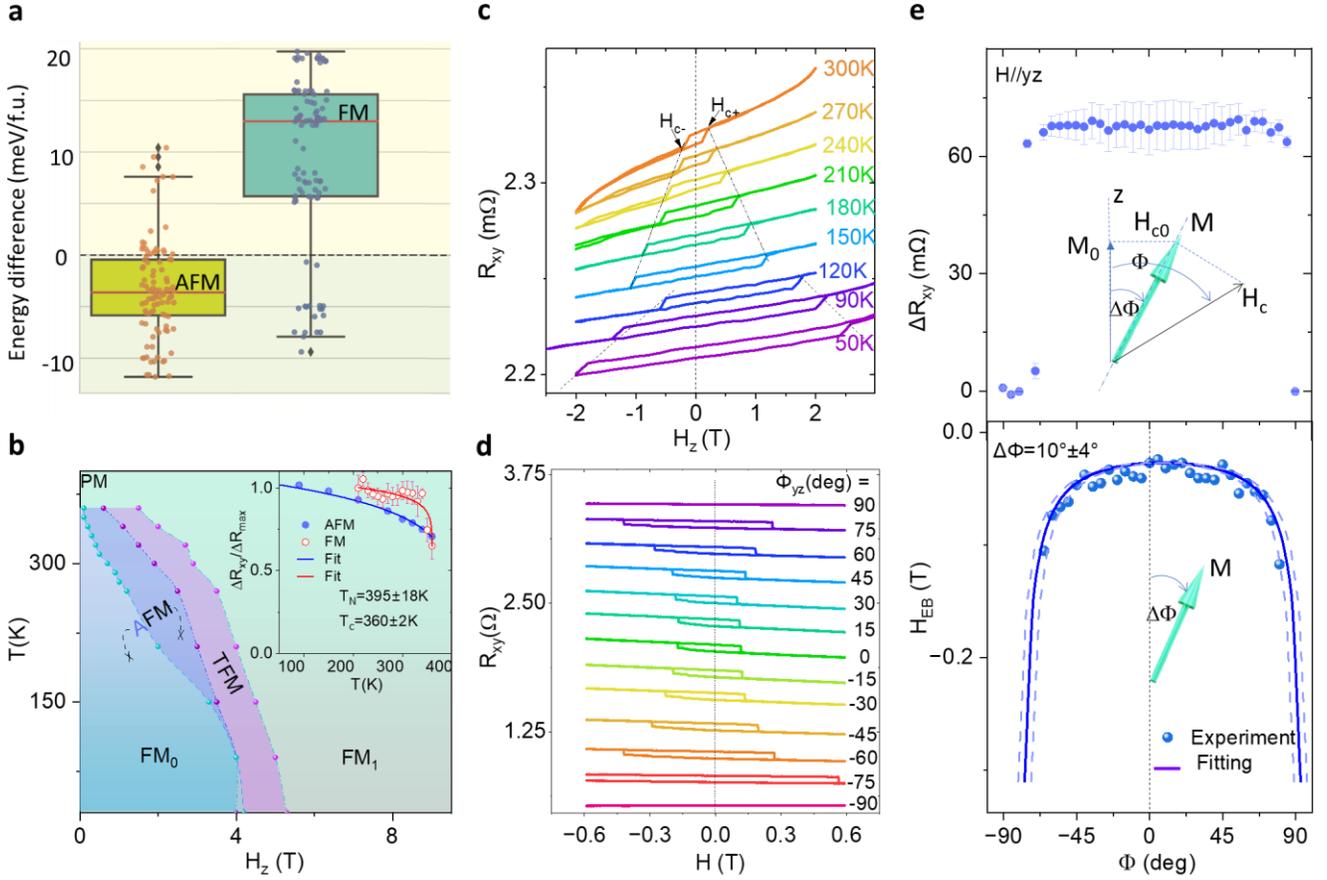

*Figure 2. Origin of the (A)FM magnetic orders and the canted magnetization M in $(Co_{0.5}Fe_{0.5})_{5-x}GeTe_2$ nanolayers. a. Boxplots of the magnetic ground state distribution of AFM (FM) orders in CFGT, where each dot represents a distinct Co distribution. The orange lines are the averaged results. b. Phase diagram of the CFGT nanolayers as a function of $H_z$ and T. The arrows illustrate the different magnetic orders. Inset shows the temperature evolution of the normalized AFM and FM components. The Curie (Néel) temperature $T_{C(N)}$ are extracted by fitting with $\sim(1-T/T_{C(N)})^\beta$. c. AHE signals $R_{xy}$ with out-of-plane field $H_z$ at different temperatures in the low field range. $H_{c+/-}$ are the coercive fields in positive/negative directions. The dashed lines are the guide to the eye. d. AHE signals $R_{xy}$ with magnetic field sweep at different yz-plane angles $\Phi$ measured at 300 K. e. The magnitude of the AHE signal $\Delta R_{xy}$ and exchange bias field $H_{EB}=(H_{c+}+H_{c-})/2$ as a function of the yz-plane angle $\Phi$. The purple curve in the bottom panel is a fit with uncertainty represented by dashed curves. The insets show the schematics of the magnetization M at a canting angle $\Delta\Phi$ and its relationship with the external field $H_c$.*

Due to the coexistence of both FM and AFM properties in CFGT nanolayers, there is a possibility that its magnetic anisotropy may be significantly impacted. To examine this, we conducted an angle-dependence study of the AHE on the CFGT Hall-bar device at room temperature (Fig. 2d). The AHE signal with exchange bias effect can be observed at all angles, except for the almost in-plane case $\Phi \approx \pm 90$ deg where the signal vanishes due to no effective field to switch the M. As shown in Fig. 2e, the magnitude of the AHE signal remains constant almost at all field sweep angles $\Phi$, indicating a strong magnetic anisotropy and minimal rotation of the magnetization towards the field direction in the range of ±600 mT. The extracted exchange bias field $H_{EB}$ at different angles is presented in Fig. 2e (bottom panel). We consider $H_{c0}$ to be the intrinsic coercive field, required to switch M with a magnetic field aligned with the easy axis of the magnetization. The projection of the nominal coercive field $H_c(\Phi)$ at field sweep angle $\Phi$ on the easy axis must be equal to $H_{c0}$, that is, $H_{c0}$ = $H_c \cos(\Phi-\Delta\Phi)$, where $\Delta\Phi$ defines the canted angle of the magnetization easy axis (Fig. 2e, insets schematics). Then the relation $H_{EB}(\Phi)=H_{EB0}/\cos(\Phi-\Delta\Phi)$ with the field rotation angle $\Phi$ can be formulated to fit the angle-dependent $H_{EB}$ (see detailed analysis in Supplementary Note 3), yielding a magnetization canting angle $\Delta\Phi$ of ~10 degrees in CFGT (see also similar results from another device in Supplementary Fig. S4). The canted magnetism is further confirmed by our DFT calculations and Monte Carlo simulations, which suggest that the intrinsic exchange bias and magnetocrystalline anisotropy energy of CFGT are the origin of the canted magnetization (see detailed analysis in Supplementary Note 4, Fig. S5 and Fig. S6).

## Field-free spin-orbit torque switching of CFGT/Pt heterostructure devices at room temperature

SOT phenomenon, which utilizes the spin-orbit interaction in materials and their hybrid structures with magnets, has emerged as an efficient method to control the magnetization in spintronic devices[3–5]. For the success of SOT-driven magnetization dynamics[3–5], it is required to enhance the SOT efficiency, achieve magnetic field-free magnetization switching, and discover new materials and methods to control the device parameters[2,6–16] (Fig. 3a). To investigate the SOT-induced magnetization switching and dynamics in CFGT samples with intrinsic exchange bias and canted magnetization effects, CFGT/Pt heterostructure Hall bar devices were nanofabricated as shown in Fig. 3b inset (see details in the Methods section). For an effective SOT magnetization switching, it is essential to have a large charge-spin conversion efficiency in Pt and high-quality CFGT/Pt interfaces for efficient transmission of spin current. In the SOT-induced magnetization switching experiment, a series of DC pulse currents $I_p$ applied along the x-direction through the spin-orbit material (Pt), generates a spin current along the z-axis with spin polarization $s_y$ along the y-axis due to the spin Hall effect. The resulting spin current exerts SOTs in CFGT (field-like $\tau_{FL}$ and damping-like $\tau_{DL}$ torques), enabling the switching of magnetization *M* direction. Specifically, the field-like torque $\tau_{FL} \sim M \times s_y$, precesses *M* about the exchange field created by spin polarization, while the damping-like torque $\tau_{DL} \sim M \times (M \times s_y)$ rotates the magnetization *M* towards the direction of spin polarization $s_y$. Typically, damping-like torque determines the switching of the magnetization[31].

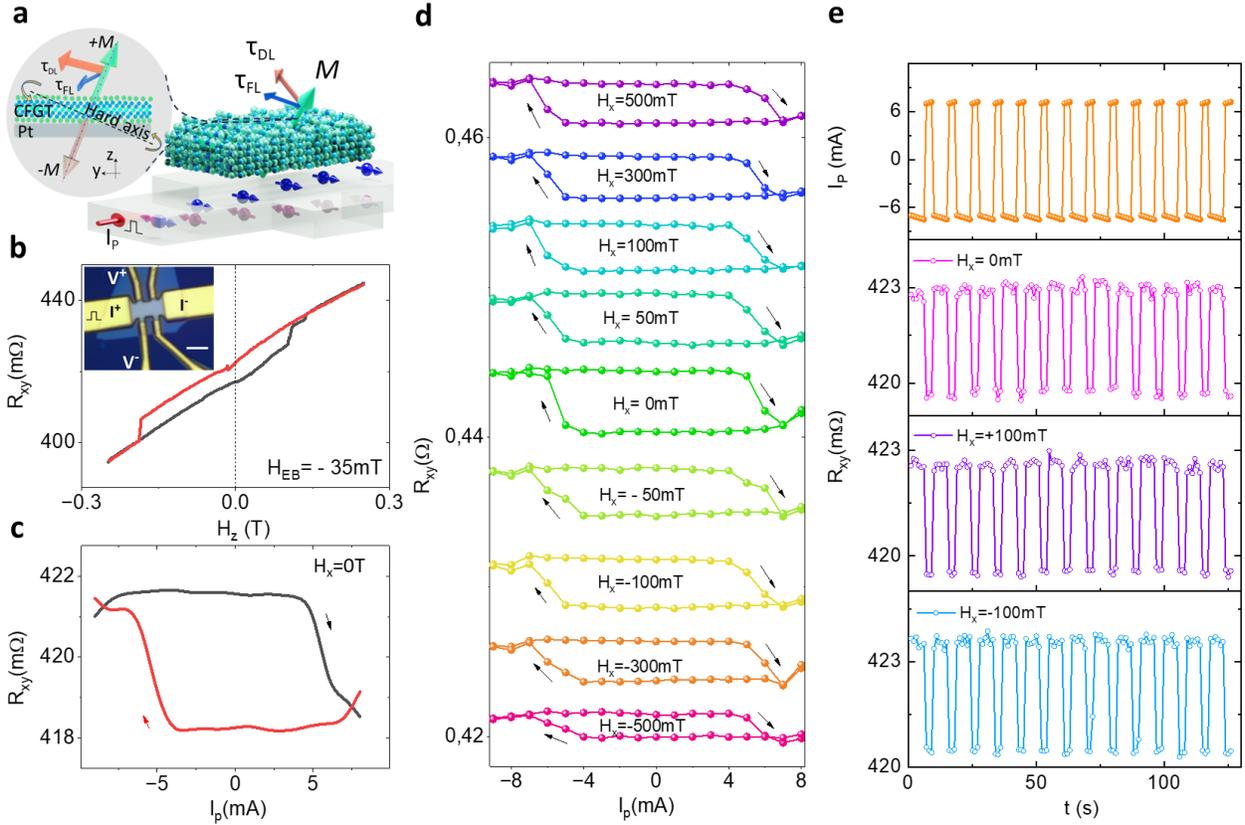

*Figure 3. Field-free spin-orbit torque magnetization switching in CFGT/Pt heterostructure at room temperature. a. Schematics of CFGT/Pt van der Waals heterostructure used for pulsed current $I_P$-induced spin-orbit torque (SOT) experiments. Inset shows the mechanism of the field-free SOT switching of CFGT. b. AHE signals with an exchange field $H_{EB}$=-35 mT. Inset shows the Optical microscope image of the CFGT/Pt heterostructure Hall-bar device. The scale bar is 2 μm. c, d. Pulsed write current $I_p$ induced transverse Hall signal $R_{xy}$ change due to SOT-induced magnetic switching without/with external in-plane magnetic field $H_x$. e. Time dependence of the pulse current $I_p$ and the corresponding measured AHE signal $R_{xy}$ with in-plane fields $H_x$=0, and ±100 mT. The pulse current dwell time is 1 ms and 3s waiting time before the read process. All the measurements here were performed in Dev 1.*

The AHE signal of the CFGT/Pt Hall device shows an out-of-plane field magnetization and the difference of the switching field for both field sweep directions suggest a strong exchange bias ($H_{EB}$= -35 mT) in CFGT (Fig. 3b). As shown in Fig. 3c, we observed the SOT-induced magnetization switching using a pulsed write current $I_p$ with pulse time 1 ms, followed by a small DC read current ($I_r$~50-500 μA) to probe the magnetization state $R_{xy}=V_{xy}/I_r$. As the signal $R_{xy}$ is proportional to the out-of-plane magnetization $M_z$[41], the SOT $R_{xy}$ signal shows current-induced magnetization change between +$M$ and -$M$. The magnitude change of $R_{xy}$ in the SOT signal is almost the same as the AHE signal with field sweep, which suggests a full switch of the CFGT magnetization. Interestingly, we observe a deterministic SOT switching of CFGT at $H_x$=0 T. This is in contrast to the conventional requirement of an in-plane magnetic field $H_x$ to break the geometrical symmetry for a deterministic switching of the magnetization with a perpendicular magnetic anisotropy (PMA) [32]. To verify the origin, we performed SOT measurements with different in-plane magnetic fields $H_x$ (Fig. 3d). The $H_x$ field does not affect the deterministic SOT switching, and the magnetization switching direction (chirality) remains the same in this field range of +/-500 mT, suggesting a robust deterministic switching against the external magnetic field due to the strong exchange bias effects and the canted magnetization of CFGT. The time-dependent switching experiments with different $H_x$ were carried out (Fig. 3e), further proving the reproducibility of the deterministic switching (also see reproducibility of SOT switching experiment Supplementary Figures S7 and discussion about the Oersted field effect on SOT switching in Supplementary Note 5).

Conventionally, chirality is determined by the external field, which favors opposite switching directions for the positive and negative in-plane magnetic fields[32]. If no external field is applied, the magnetization would be pulled to the in-plane orientation and broken into multiple domains, instead of the deterministic switching. However, in the canted magnetization scenario of the CFGT sample (Fig. 3a, inset), the symmetry is broken as the easy axis of the magnetization *M* is canted at an angle ΔΦ with respect to the normal direction. The damping-like torque $τ_{DL}$ can rotate the *M* anticlockwise over the hard axis towards another easy axis, while the other rotation direction is not possible. This also explains the unchanged chirality of SOT switching with the external field direction, which is similar to that observed in other canted magnets[33]. Thermal fluctuations induced by pulse current can randomize the magnetic domain of CFGT and make switching behavior unstable[34]. This is not the case in our CFGT/Pt SOT device, although the reduction of the SOT signal at a larger current due to thermal fluctuations cannot be ruled out[35]. We can also rule out thermally assisted switching if the device reaches above $T_c$, as this can result in non-deterministic switching due to the formation of multiple domains[36,37]. Therefore, the observed deterministic and field-free switching of CFGT can be attributed to the dominant SOT contribution and the canted magnetization of CFGT.

**Estimation of spin-orbit torque magnetization dynamics using 2$^{nd}$ harmonic Hall measurement in CFGT/Pt heterostructure devices**

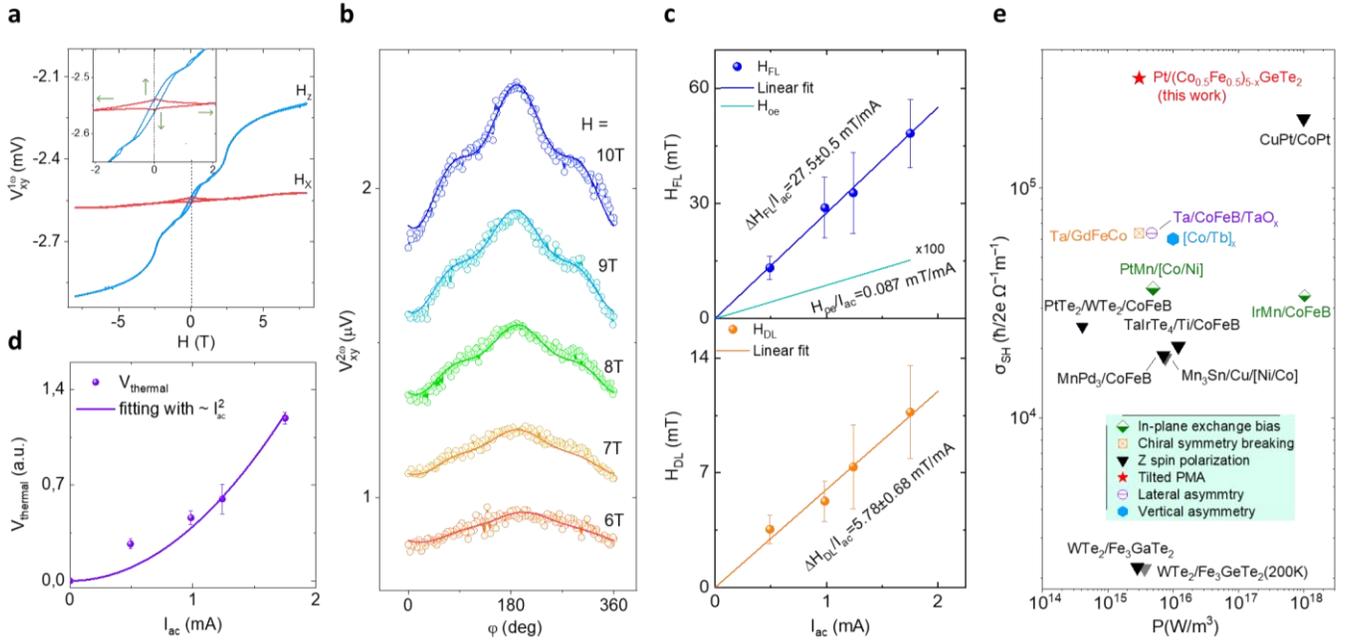

*Figure 4. Harmonic Hall measurements of the CFGT/Pt Hall-bar device at room temperature. **a.** $1^{st}$ harmonic $V_{xy}^{1\omega}$ signals as a function of the out-of-plane field $H_z$ and in-plane field $H_x$ sweeps. The inset is the zoom-in around 0 T. The green arrows show the FM components. **b.** In-plane angle φ dependence of the $2^{nd}$ Harmonic signal $V_{xy}^{2\omega}$ at $I_{ac}$=1.75 mA with in-plane field H. **c, d.** Extracted current dependence of the field-like effective field $H_{FL}$, Oersted field $H_{Oe}$, damping-like effective field $H_{DL}$, and thermal contribution. All the harmonic measurements were performed with ω=213.14 Hz at 300 K in Dev 2. **e.** Benchmarking our Pt-CFGT results with the state-of-the-art field-free magnetization switching devices reported in the literature such as PtMn/[Co/Ni][9], IrMn/CoFeB[10], Ta/GaFeCo[38], CuPt/CoPt[39], TaIrTe$_4$/Ti/CoFeB[8], PtTe$_2$/WTe$_2$/CoFeB[40], MnPd$_3$/CoFeB[41], Mn$_3$Sn/Cu/[Ni/Co][42], WTe$_2$/Fe$_3$GaTe$_2$[43], WTe$_2$/Fe$_3$GeTe$_2$[6], Ta/CoFeB/TaOx[11], [Co/Tb]$_x$[44]. Spin Hall conductivity $\sigma_{SH}$ as a function of the dissipation power density for magnetization switching P=($J_{sw}^2/\sigma_c$), where $J_{sw}$ is the switching current density and $\sigma_c$ is the conductance of the spin source material. Different field-free mechanisms are indicated with the symbols.*

Harmonic Hall measurement is a powerful tool for studying different SOT contributions and effective spin-orbit fields in magnetic heterostructures[45]. Application of an AC current $I_{ac}$ = $I_0$sinωt with frequency ω modulates the SOT amplitude and induces small oscillations of M about its equilibrium direction[44]. Such oscillations generate a $2^{nd}$ harmonic contribution to the Hall voltage[46] ($V_{xy}$≈$V_0$+$V_{xy}^{1\omega}$sinωt+$V_{xy}^{2\omega}$cos2ωt), where $V_{xy}$ depends on $M_z$ through the AHE and the product of $M_x$ x $M_y$ through the planar Hall effect (PHE). The $1^{st}$ harmonic term $V_{xy}^{1\omega}$ relates to the equilibrium direction of the magnetization and is independent of the modulated SOT, which was measured as an in-phase (Ø$_{lockin}$=0°) signal. In the $1^{st}$ harmonic Hall signals $V_{xy}^{1\omega}$ vs. $H_z$ field sweep (Fig. 4a), we observed the expected AFM phase transition and FM hysteresis in high and low field ranges, respectively, whereas the $H_x$ field sweep measurement shows the FM magnetization component being pulled to the in-plane direction at around 2 T (i.e. magnetic anisotropy field, $H_k$), without any clear indication of the AFM transition up to 7 T.

The out-of-phase (Ø$_{lockin}$=90°) 2$^{nd}$ harmonic term $V_{xy}^{2\omega}$ measures the response of the magnetization to the current-induced spin-orbit fields. The current-induced effective field H$_{eff}$ can be decomposed into two components: the field-like (H$_{FL}$) and damping-like (H$_{DL}$) spin-orbit fields, which can be extracted by the following equation[45],

$$V_{xy}^{2\omega} = -(V_{AHE}\frac{H_{DL}}{H-H_k} + V_{thermal})\cos\varphi + 2V_{PHE}\frac{H_{FL}+H_{oe}}{H}(2\cos^3\varphi - \cos\varphi); \quad \text{(Eq. 1)}$$

where H is the applied external field; $\varphi$ is the angle between the applied current and the field. V$_{AHE}$ and V$_{PHE}$ are the AHE voltage and planar Hall (PHE) voltage, respectively. V$_{thermal}$ is the thermal-related contribution, like anomalous Nernst effect and spin Seebeck effect, etc. By adopting the simplified model[47], the current induced Oersted field H$_{oe}$= I$_{Pt}$/(2W) is also calculated, where W is the width of the Hall device. Therefore, the 2$^{nd}$ harmonic Hall voltage signal $V_{xy}^{2\omega}$ as a function of an in-plane rotation $\varphi$ with fixed magnetic field H can be measured to extract the contributions from H$_{DL}$, H$_{FL}$, and the current-induced thermal effect (Fig. 4b). Furthermore, the current dependence of the contributions from H$_{DL}$, H$_{FL}$, and the thermal effects are plotted in Fig. 4c and Fig. 4d. We extracted ΔH$_{DL}$/J$_{ac}$=4.2 mT per MA/cm$^2$ with an effective saturation magnetization µ$_0$M$_s$ ≈ 5.9 mT and ΔH$_{FL}$/J$_{ac}$=19.7 mT per MA/cm$^2$ (see more details on the harmonic Hall measurement methods in Supplementary Note 6, Supplementary Fig. S8 and the summary of the key SOT device parameters in Supplementary Table S2). The µ$_0$M$_s$ is much smaller than the value of pure FM magnetic phase with the AA' crystal structure[30]. This suggests that the AFM phase does not contribute much to the effective saturation magnetization even at large in-plane fields but coexists with the FM phase, which agrees with the 1$^{st}$ harmonic result with H$_x$ sweep.

The current-induced Oersted field is negligible and cannot be the origin of the field-free SOT-induced magnetization switching, while thermal contribution to the harmonic Hall voltage signal $V_{xy,thermal} \propto I^2$, increasing quadratically with the current, is proportional to $M_x$. Geometrically, $V_{xy,thermal} \sim M_x \times \nabla T_z$, where $\nabla T_z$ is due to the different thermal conductivity of CFGT, Pt, SiO$_2$/Si substrate, and air. Utilizing both SOT-induced magnetic switching and 2$^{nd}$ harmonic Hall measurements, we can conclude that the canted magnetization in CFGT can be efficiently controlled with a switching current density of J$_{sw}$~8 x10$^6$ A/cm$^2$ at 300 K, which is lower than the reported values for vdW magnets at low temperatures[48]. Moreover, as benchmarked against the state-of-the-art representative field-free SOT devices (Fig. 4e), including devices with lateral asymmetry (mirror symmetry breaking), chiral symmetry breaking, in-plane exchange bias, vertical asymmetry, out-of-plane spin polarization $S_z$, Our Pt/CFGT devices show a large charge-spin conversion efficiency and low operating power density. Pt has been widely used as a spin-orbit material as it has a large spin Hall conductivity $\sigma_{s,y} \sim 10^6\ \hbar/2e\ \Omega^{-1}\ m^{-1}$ [34,48,49] and has also been used together with vdW magnets with strong PMA. However, SOT switching demonstrated so far in Pt/vdW-magnet with PMA needs an assisted external magnetic field for deterministic switching (see state-of-the-art vdW magnet-based SOT devices in Supplementary Note 7 and Supplementary Fig. S9). Alternatively, materials with low crystal symmetry, such as $WTe_2$ and $TaIrTe_4$ with an out-of-plane SOT component to deterministically switch the magnetization without an external field[6–8,13,50]. Unfortunately, the unconventional SOT efficiency of $WTe_2$ and $TaIrTe_4$ is still much smaller than the $\sigma_{s,y}$ of Pt, giving rise to higher power consumption[6–8,50].

**Conclusion and Perspective**

We observe the coexisting FM and AFM properties in vdW magnet CFGT much above room temperature, with intrinsic exchange bias and canted perpendicular ferromagnetism, which is not possible to attain in conventional systems. Such intrinsic coexistence of non-trivial magnetic orders enables the demonstration of an energy-efficient, field-free, and deterministic switching of the vdW magnet CFGT in heterostructure with conventional material Pt with a large spin Hall conductivity. The detailed investigation of magnetic properties of CFGT and magnetization dynamics using SOT switching and $2^{nd}$ harmonic Hall measurements of CFGT/Pt heterostructures show their relationship with the observed field-free SOT behavior. By demonstrating the feasibility of deterministic SOT-induced magnetization switching in a non-trivial vdW magnet at room temperature without the application of an external magnetic field, our work lays the foundation for the development of efficient spintronic devices based on vdW magnets. Moreover, intrinsically CFGT hosts both FM and AFM magnetic orders with a canted out-of-plane magnetism, offering the flexibility to combine other materials with even larger charge-spin conversion efficiency to enhance the device's performance.

**Methods**

**Device Fabrication:** Single crystals of $(Fe_{0.5}Co_{0.5})_5GeTe_2$ (CFGT) were grown by chemical vapor transfer (CVT) method by HqGraphene. The AHE measurements were measured in Hall-bar devices of CFGT (15-60 nm) on $SiO_2$/Si substrate. The CFGT crystal was exfoliated onto the $SiO_2$/Si wafer, followed by electron-beam lithography (EBL) and Ar plasma etching to pattern in a Hall-bar geometry, and the Au/Ti contacts were deposited $2^{nd}$ EBL step and lift-off technique. The CFGT/Pt heterostructures were prepared by exfoliating thin CFGT crystals (15-60 nm) inside an $N_2$ glovebox onto pre-deposited Pt(10 nm)/Ti(2 nm) layers on a $SiO_2$/Si substrate. The Pt layer was prepared by electron-beam evaporation onto a Si/$SiO_2$ substrate with a Ti seed layer. After exfoliation of CFGT, the samples were immediately transferred to an electron-beam evaporation system to deposit a 2 nm Al layer, forming an AlO$_x$ capping layer. The Hall bar and contacts were patterned using EBL and Ar ion beam etching.

**Electrical measurements:** Low temperature and high magnetic field measurements of CFGT Hall bar devices were performed in the Quantum Design cryogen-free PPMS DynaCool system with bias current in the range of 50-500 µA in the temperature range of 10-360 K and magnetic field up to 13 Tesla (T).

**SOT switching and Harmonic measurements** in a low magnetic field range were carried out in a vacuum cryostat with a magnetic field up to 0.7 T for CFGT/Pt devices. The electronic measurements were performed using a Keithley 6221 current source, and a nanovoltmeter 2182A. To monitor the longitudinal and transverse Hall resistances, Keithley 2182A nanovoltmeters were utilized. For SOT-induced magnetization switching measurements, Keithley 2182A nanovoltmeters were used to monitor the response of the Hall resistances, while a Keithley 6221 AC source was utilized with a pulse current of 1 millisecond (ms) through the device, followed by a DC read current. The harmonic measurement was performed using Lockin SR830 to measure in-phase 1$^{st}$ and out-of-phase 2$^{nd}$ harmonic voltages, respectively. The 2$^{nd}$ harmonic measurements in the high magnetic field range were carried out in the Quantum Design cryogen-free PPMS DynaCool system with an external electronic connection to Lockin SR830 to measure the 1$^{st}$ and 2$^{nd}$ harmonic voltages.

**SQUID measurements:** A Quantum Design superconducting quantum interference device (SQUID) was used to measure the static magnetic properties of bulk CFGT crystals. We measured the magnetization as a function of the out-of-plane (H‖z) magnetic field. The bulk CFGT crystal was glued on a Si substrate to properly align the sample in the magnetic field during the measurements.

**STEM measurements:** Scanning transmission electron microscopy (STEM) measurements were carried out using a JEOL monochromatic ARM200F transmission electron microscope (TEM), which is equipped with a Schottky field emission gun, a double-Wien monochromator, a probe Cs corrector and an image Cs corrector, as well as high angle annular dark field (HAADF) detectors for STEM imaging, a double silicon drift detector (SDD) for energy dispersive X-ray spectroscopy (EDXS), and a Gatan image filter (GIF) Continuum. The microscope was operated at 200 kV for HAADF STEM imaging. The beam convergence half-angle and inner collection half-angle for HAADF STEM are ~ 27 mrad and ~ 55 mrad, respectively. Specimens for STEM measurements were prepared by using a FEI Versa 3D focus ion beam–scanning electron microscope (FIB-SEM). After depositing a protection layer containing Pt and C using first the electron beam and then the Ga ion beam in the FIB-SEM, a lamella of the material was cut out using an ion beam at 30 kV and 1 nA. After transferring the lamella to a Cu TEM grid, the lamella was gradually thinned down by the ion beam. The thinning process was performed first at 30 kV and with a gradually decreasing beam current from 1 nA to 100 pA. Then, the gentle polishing of the specimen was carried out with an ion beam energy of 5 kV and 2 kV to minimize the ion beam effect.

**Acknowledgements**

Authors acknowledge funding from European Union (EU) Graphene Flagship project 2DSPIN-TECH (No. 101135853), 2D TECH VINNOVA competence center (No. 2019-00068), Wallenberg Initiative Materials Science for Sustainability (WISE) funded by the Knut and Alice Wallenberg Foundation, EU Graphene Flagship (Core 3, No. 881603), Swedish Research Council (VR) grant (No. 2021–04821, No. 2018-07046), FLAG-ERA project 2DSOTECH (VR No. 2021-05925) and MagicTune, Graphene Center, Chalmers-Max IV collaboration grant, Areas of Advance (AoA) Nano, AoA Materials Science and AoA Energy programs at Chalmers University of Technology. We also acknowledge the help of staff at the Quantum Device Physics and Nanofabrication laboratory at Chalmers University of Technology. L.Z and E.O. acknowledge the Swedish Research Council and Swedish Foundation for Strategic Research for access to ARTEMI, the Swedish National Infrastructure in Advanced Electron Microscopy (2021-00171 and RIF21-0026). This work was performed in part at the Chalmers Material Analysis Laboratory, CMAL. B.S. acknowledges financial support from the Swedish Research Council (grant no. 2022-04309 and grant No. 2018-07082). The computations were enabled by resources provided by the National Academic Infrastructure for Supercomputing in Sweden (NAISS) at UPPMAX (NAISS 2023/5-238) and at NSC and PDC (NAISS 2023/3-42) partially funded by the Swedish Research Council through grant agreement no. 2022-06725. B.S. also acknowledges the allocation of supercomputing hours in the EuroHPC Development Access Call (EHPC-DEV-2024D04-071) to EuroHPC resources in LUMI-C in Finland.

**Data availability**

The data that support the findings of this study are available from the corresponding authors at a reasonable request.

**Contributions**

B.Z. and S.P.D. conceived the idea and designed the experiments. B.Z. fabricated and characterized the devices with support from L.B. and R.N.. S.E. and B.S. performed the theoretical calculations and data analysis. L.Z and E.O. performed STEM characterization and data analysis. P.S. performed SQUID magnetization measurements and data analysis. B.Z., S.E. B.S, and S.P.D. analyzed and interpreted the data, compiled the figures, and wrote the manuscript with feedback from all co-authors. S.P.D. supervised and managed the research project.

**Corresponding authors**

Correspondence to Bing Zhao, zbing@chalmers.se; Biplab Sanyal, biplab.sanyal@physics.uu.se; Saroj P. Dash, saroj.dash@chalmers.se

**Competing interests**

The authors declare no competing interests.